\def\year{2020}\relax
\documentclass[letterpaper]{article} 
\usepackage{aaai20}  
\usepackage{times}  
\usepackage{helvet} 
\usepackage{courier}  
\usepackage[hyphens]{url}  
\usepackage{graphicx} 
\usepackage{graphicx}  
\usepackage[capitalize,noabbrev,nameinlink]{cleveref}  
\newcommand{\system}{EventMapper }
\newcommand{\PP}[1]{
	\vspace{2px}
	\noindent{\bf {#1}{.}}
}

\newcommand{\squishitemize}{
\begin{list}{$\bullet$}
	{ \setlength{\itemsep}{0pt}
		\setlength{\parsep}{3pt}
		\setlength{\topsep}{3pt}
		\setlength{\partopsep}{0pt}
		\setlength{\leftmargin}{1.95em}
		\setlength{\labelwidth}{1.5em}
		\setlength{\labelsep}{0.5em} } }

\newcounter{Lcount}
\newcommand{\squishlist}{
	\begin{list}{\arabic{Lcount}. }
		{ \usecounter{Lcount}
			\setlength{\itemsep}{0pt}
			\setlength{\parsep}{3pt}
			\setlength{\topsep}{3pt}
			\setlength{\partopsep}{0pt}
			\setlength{\leftmargin}{2em}
			\setlength{\labelwidth}{1.5em}
			\setlength{\labelsep}{0.5em} } }
	
\newcommand{\squishend}{\end{list}}

\title{EventMapper: Detecting Real-World Physical Events Using Corroborative and Probabilistic Sources}
\author{
Abhijit Suprem and Calton Pu\\
School of Computer Science\\ 
Georgia Institute of Technology,\\
Atlanta, GA\\
asuprem@gatech.edu 
}
 
\begin{document}

\maketitle

\begin{abstract}
	
The ubiquity of social media makes it a rich source for physical event detection, such as disasters, and as a potential resource for crisis management resource allocation. There have been some recent works on leveraging social media sources for retrospective, after-the-fact event detection of large events such as earthquakes or hurricanes. Similarly, there is a long history of using traditional physical sensors such as climate satellites to perform regional event detection. However, combining social media with corroborative physical sensors for real-time, accurate, and global physical detection has remained unexplored.
	
This paper presents EventMapper, a framework to support event recognition of small yet equally costly events (landslides, flooding, wildfires). EventMapper integrates high-latency, high-accuracy corroborative sources such as physical sensors with low-latency, noisy probabilistic sources such as social media streams to deliver real-time, global event recognition. Furthermore, EventMapper is resilient to the concept drift phenomenon, where machine learning models require continuous fine-tuning to maintain high performance. 
	
By exploiting the common features of probabilistic and corroborative sources, EventMapper automates machine learning model updates, maintenance, and fine-tuning. We describe three applications built on EventMapper for landslide, wildfire, and flooding detection.

\end{abstract}

\section{Introduction}
\begin{figure}[t]
	\centering
	\includegraphics[width=1\linewidth]{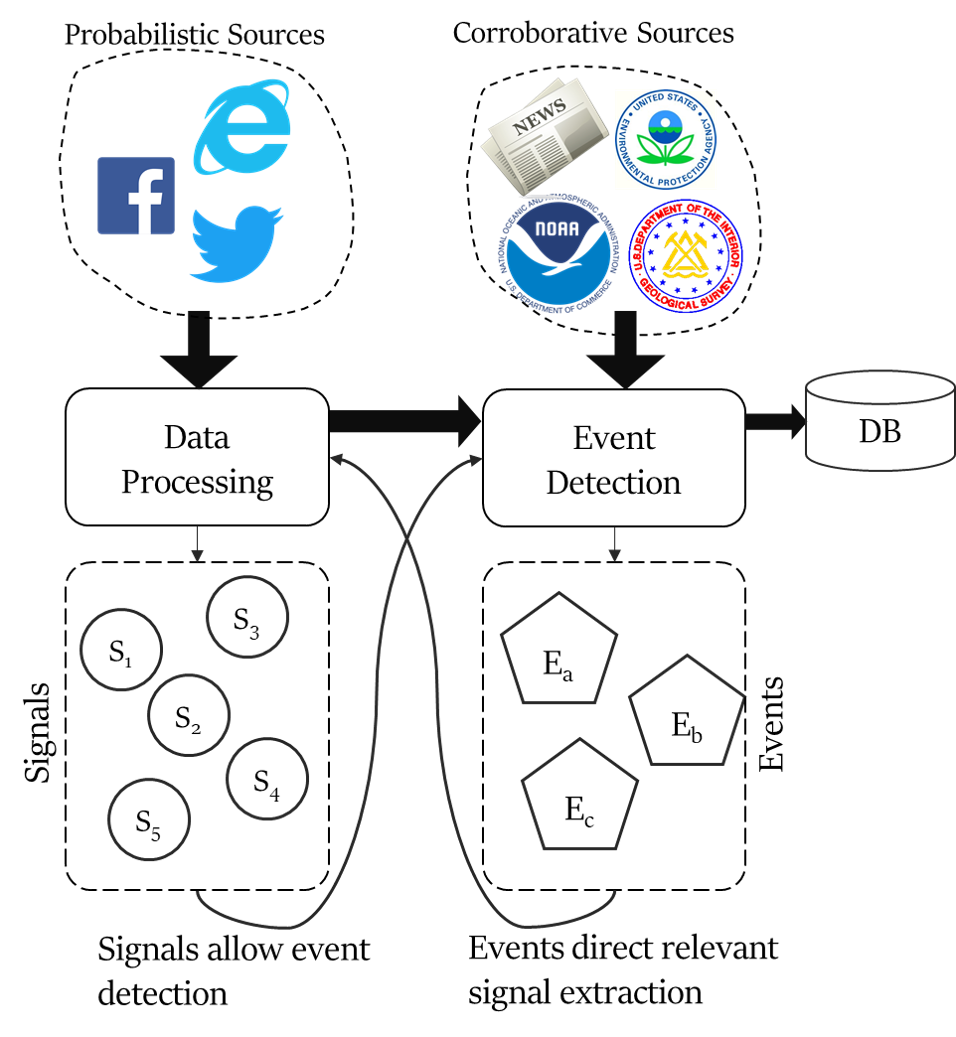}
	\caption{\textbf{Co-dependence of Data Processing and Event Detection.} Data Processing identifies relevant signals in probabilistic sources. Event Detection on corroborative sources fine-tunes Data Pro-cessing for robustness against drift}
	\label{fig:codependence}
\end{figure}

Event recognition, which is the classification and re-identification of relevant events over time, has a long history \cite{cepsummary,sakaki2010earthquake,hurricane,gft,assed,litmus}. Event recognition comprises of two intertwined processes: (i) data processing and (ii) event detection, that have co-dependence: event detection requires processing raw real-world data to extract relevant signals, and useful data processing requires knowing which events to follow in the universe of events (\cref{fig:codependence}). 

There have been various works on data processing, event detection, and complex event detection \cite{cepsummary}; furthermore, emergence of powerful compute resources combined with large volume streaming data has led to recent works on event detection with stream processing \cite{stream} and machine learning \cite{assed}. More recently, there have been works on exploiting human sensors from social media sources for dense global physical event recognition, such as earthquake \cite{sakaki2010earthquake}, hurricane \cite{hurricane}, and landslide detection \cite{litmus}.

\begin{figure}[t]
	\centering
	\includegraphics[width=1\linewidth]{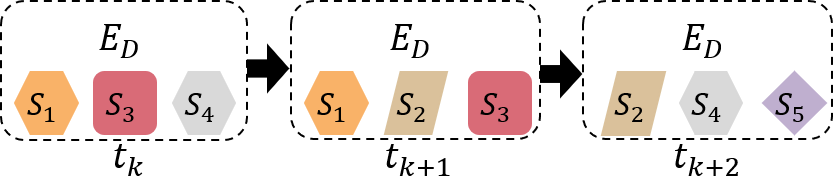}
	\caption{\textbf{Drift in signals for an event over time.} The signals composing an event change due to different types of concept drift. For example, due to cyclical drift, Signal 4 disappears at $t_{k+1}$ and reappears at $t_{k+2}$.}
	\label{fig:eventdrift}
\end{figure}

A common challenge in event recognition is the concept drift phenomenon, where the distribution of data changes continuously. The concept drift phenomenon is well-documented in event recognition \cite{gamaA,gamaB}. Under concept drift, static event detection methods (rule-based, machine learning, or deep learning methods) exhibit performance degradation over time. So far, effective concept drift adaptation strategies for noisy social media sources data remains unexplored; most work in drift detection and adaptation focuses on closed, small datasets with well known drift attributes \cite{openset}.

In this paper, we present \system, a framework for event recognition that exploits the co-dependence of data processing and event detection to support long-term event recognition that can adapt to concept drift in social media sources. \system integrates two types of sources: high-confidence corroborative sources and low-confidence probabilistic sources.

\PP{Corroborative Sources} A high-confidence corroborative source is a dedicated physical or web-based sensor that provides annotated physical event information based on experts. Due to expert corroboration, corroborative sources have reduced noise and drift \cite{coolpaper}. However, corroboration increases their cost; so corroborative sources have reduced coverage and higher latency.

\PP{Probabilistic Sources} A probabilistic source is any source without corroboration, such as raw web streams or human sensors. Lack of corroboration makes such streams, such as Twitter and Facebook, noisy and drifting \cite{noisy}. However, such sources are globally available and have low latency \cite{coolpaper}.

The \system framework allows deployment of systems for weak-signal event recognition. In contrast to strong-signal events such as earthquakes and hurricanes, which have hundreds of thousands of corroborative and probabilistic signals per event, weak-signal events have 1-3 signals per event. Strong-signal events have easily separable signals and detection can be per-formed retrospectively with trend analysis \cite{sakaki2010earthquake,hurricane}. Conversely, weak-signal event detection is more difficult since signals are not easily separable. Since each event has few signals, they are hidden in noise and drift and require precise data processing with statistical and machine learning methods.

By integrating corroborative and probabilistic sources by exploiting their co-dependence, \system improves upon static event recognition for weak-signal events: (i) corroborative sources, which are used directly for event detection, fine-tune the data processing modules for social sensors; consequently, data processing modules remain robust to concept drift over time, and (ii) data processing modules use statistical and machine learning methods to extract relevant signals from probabilistic sources for event detection.

Our contributions are as follows:

\squishlist
\item We present the \system framework; we will be releasing the code to the open-source community. The \system framework supports weak-signal event detection while remaining ro-bust to concept drift. By exploiting the co-dependence between data processing and event detection, \system automates concept drift adaptation. 
\item We describe three applications deployed on the \system framework for three weak-signal events: landslide detection, wildfire detection, and flooding detection. For each application, we show evidence of that weak-signal nature of the events. We will provide links to the application demos.
\squishend

\section{Preliminaries}

\begin{figure}[t]
	\centering
	\includegraphics[width=1\linewidth]{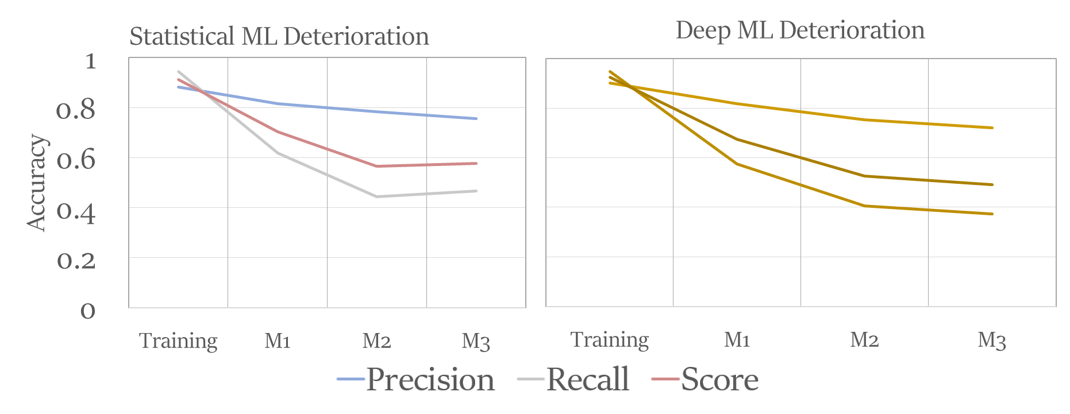}
	\caption{\textbf{ML Classifier deterioration} We compare statistical and deep models over time on our data (described in \textbf{Applications} section). After initial training, models are evaluated on subsequent months (M1, M2, M3).}
	\label{fig:deterioration}
\end{figure}

\subsection{Events and Signals} 

The concept drift phenomenon is well known in event detection \cite{gamaA,gamaB}. Under concept drift, the signals characterizing an event change over time as new signals are introduced. Also, existing signals may disappear. With desired event $E_D\in E$ in $S$, the social media stream, we aim to detect it from stream elements, e.g. a post written in natural language and decomposed into numeric signals $s$  with w2v \cite{w2v} for event detection. When new words enter the vocabulary, new signals are introduced. Furthermore, recent approaches in NLP use surrounding context of words to create a word context vector \cite{context}; as a word's definition changes due to memes or viral content, the context vector also changes.

We characterize concept drift in terms of events and signals. Each post $S_p$ in the stream $S$ is a distribution over the events $P(E|S_p)$, which includes $E_D$. Each post is also a generative model over the component signals $P(S_p|s)$. So, 

$$E_D=\sum_i^k a_is_i$$

where $k$ is the number of signals present in the stream; $k$ changes as new words are added, and existing words become obsolete. Changes in word meanings also change $k$. $a_i$ is the strength of a signal $s_i$ in the desired event.

\subsection{Concept Drift} 

Concept drift occurs when the distribution of $a_i$ and $s_i$ change for a desired event $E_D$ (\cref{fig:eventdrift}). Since an ML classifier learns a projection from the signals domain to the classification domain (e.g. binary classification), changes in the signals domain due to drift makes the learned projection invalid. Under drift, a static ML classifier will exhibit performance degradation over time. We show this in \cref{fig:deterioration}, where we used static classifiers trained on landslide data from 2014 (see \textbf{Implementation} section on data collection steps) to evaluate performance on 2017, 2018, and 2019. In each case, performance degrades over time due to drift in the online data stream and increasing social media noise that renders older models obsolete. Several variants of drift are known: (i) gradual drift slowly changes the coefficients of existing signals and adds new signals, with lexical diffusion \cite{lexical} as a representative example; (ii) in contrast, sudden drift causes rapid changes in coefficients, with viral memes as a representative example; (iii) cyclic drift re-introduces signals periodically before disappearing them, e.g. with landslide detection, our system needs to be aware of election landslide related posts each October and November in the US; and (iv) flash drift introduces new meanings for a short time.

\section{Related Work}

\begin{figure}[t]
	\centering
	\includegraphics[width=1\linewidth]{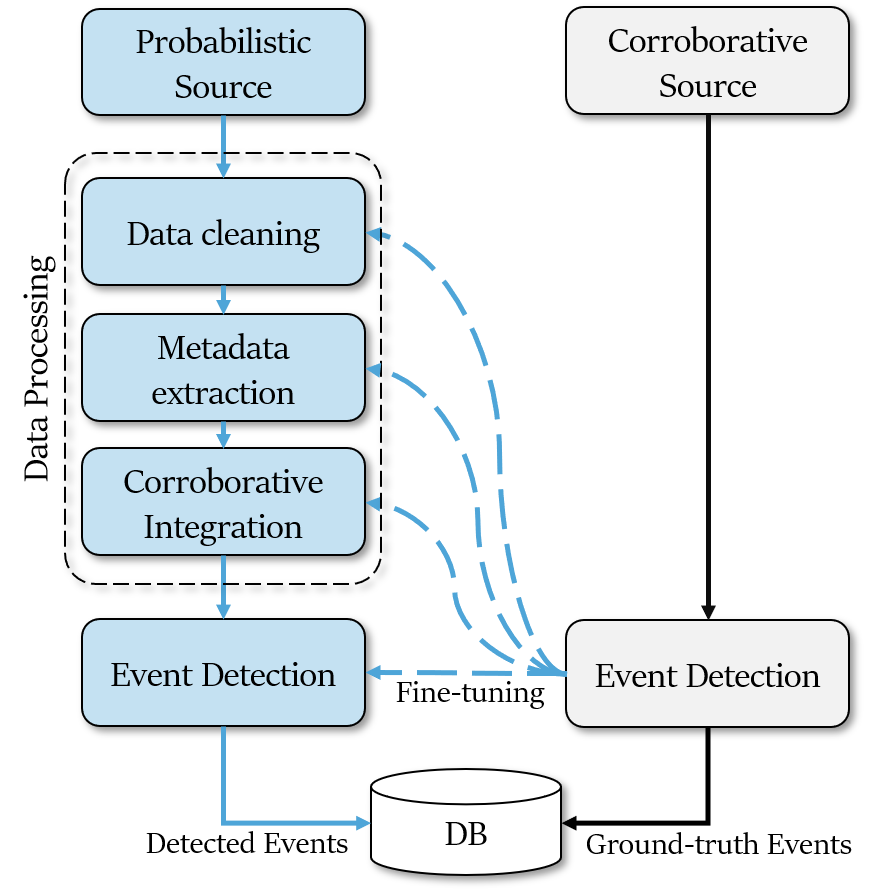}
	\caption{\textbf{\system} The \system dataflow integrates corroborative and probabilistic sources; corroborative sources provide ground truth event detection, and probabilistic sources provide real-time, global coverage.}
	\label{fig:system}
\end{figure}

\label{sec:related}
Some challenges particular to the streaming domain include: difficulty of concept drift, natural language processing (NLP) on short text data, and weak-signal event detection. We introduce related work for each area and tie them back to event recognition in the streaming domain.

\subsection{Concept Drift} 	

Recent works have focused on adapting the static classifiers to the dynamic, streaming domain. However, these approaches keep many of the data assumptions of the static models; a thorough survey is available in \cite{gamaA,gamaB}. 

\PP{Assumptions} Under the \textbf{closed data} assumption, streaming data is well specified by the training data and large amounts of ground truth labels can be quickly generated for model updates. Under the \textbf{immediate feedback} assumption, oracle labels are available for drift updates; recent approaches have relied on weak supervision in lieu of oracles \cite{weaksupervision}.

The approach in \cite{kme} uses several drift detector modules combined with updates to create an adaptive ensemble. Windowing is used in \cite{windowing} to  track concept drift and record irrelevant data to retroactively fix prior cloassifications. The KNN-based approach in \cite{samknn} uses nearest neighbor search to identify the best window of models for a sample. Finally, \cite{gamaB,EDDM} use explicit error tracking to detect drift.

Concept drift in online streams has seen limited investigation. Recent works include \cite{cepdrift}, which explore complex event detection in the presence of sensor drift. Further, \cite{lexical} explore lexical diffusion, an example of gradual drift in signals and coefficients based on geographical location. 

Many of the existing approaches in event detection also assume data without concept drift. Such assumptions, which were made in Google Flu Trends (GFT),  create models that degrade over time. GFT was originally created to complement the CDCs flu tracking efforts by identifying seasonal trends in the flu season \cite{gft}. Failure to account for seasonal changes in event characteristics led to increasing errors over the years, and by 2013, GFT missed the trends by 140\%. This error has been attributed to exclusion of new data from CDC, changes in the underlying search data distribution itself, and cyclical data artifacts \cite{gftfail1,gftfail2,gftfail3}.

\PP{Short Text Streams} NLP plays a key role in extracting useful signals  from  text. However, NLP is suited for  long-text data; since social media text is short and noise \cite{noisy}, traditional NLP techniques are not sufficient for classification \cite{shorttext}.

\PP{Weak-Signal Event Detection} Event recognition on web streams have primarily focused on strong-signal events, such as earthquakes \cite{sakaki2010earthquake}, flu \cite{gft}, and hurricans \cite{hurricane}. These are large signal events since cases can be verified and have abundant reputable data. We focus on weak-signal events that have little to no corroboration (for our events detected on social media, only 5\% of events have corroboration from reputable sources; the rest are classified using our system). Specifically, we focus on landslides, wildfires, and flooding detection.

\section{The EventMapper Framework}
\label{sec:system}

We now describe our \system  framework. We first cover the framework dataflow at a high level. We then cover the probabilistic source event detection that is integral to \system. Finally, we provide salient implementation details.

\subsection{\system Dataflow}
The EventMapper framework, shown in \cref{fig:system}, integrates corroborative and probabilistic sources for dense, global, real-time physical event recognition. Ground truth events detected from corroborative sources are used to fine-tune data processing (which includes data cleaning, metadata extraction, and ML models for classification) for probabilistic sources. Continuous fine-tuning requires concept drift adaptation, which means updating data processing modules with the current stream's distribution. Current approaches described in Related Work perform this update manually and in the closed dataset domain. \system's advantage is in automating the continuous fine-tuning, allowing scalable drift adaptation that remains functional long after initial model construction. This allows the data processing steps to remain robust to concept drift.

\begin{figure}[t]
	\centering
	\includegraphics[width=1\linewidth]{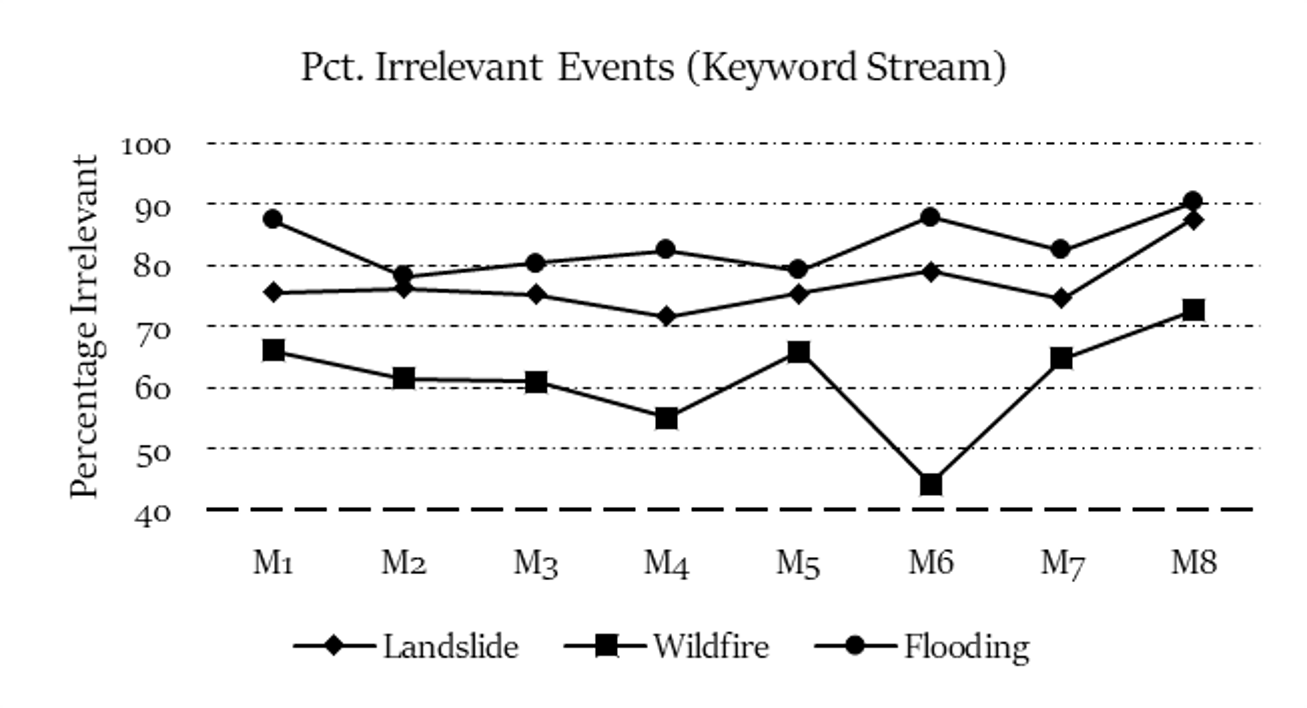}
	\caption{\textbf{Irrelevant Events} Percentage irrelevant events for each event type (landslide, wildfire, flooding). Over 60\% events from keyword stream are irrelevant to event type. These irrelevant items remain after  we use keyword filtering (e.g. using the \textit{landslide} keyword) to download social media posts and  static stopwords to filter out some items (e.g. \textit{election} to filter out election results).}
	\label{fig:irrelevant}
\end{figure}

\PP{Corroborative Source} The corroborative sources provide ground truth events. Since corroborative sources use multiple sources and human experts, they are slower and have low coverage. For example, news coverage about landslides appear 2-3 days after the event has occurred, which renders event detection based on a news article irrelevant since the delay between event and detection is too great. News articles also do not have dense global coverage, since smaller landslides in rural areas may not be reported.

\PP{Probabilistic Source} The probabilistic sources are derived from social media sources. Since probabilistic sources represent a large variety of events, data processing is required to identify relevant signals for event detection. In contrast to corroborative sources, where each source represents a specific event (e.g. NASA MODIS detects wildfires only), a probabilistic source such as a Twitter stream covers multiple events. Even with keyword-based streams, the presence of lexical diffusion, multiple word meanings (polysemy) and memes increase noise. For example, landslide detection on Twitter by following tweets using the word landslide or mudslide also returns tweets for election landslides, the ice cream Mississippi Mudslide, and the song \textit{Landslide} by Fleetwood Mac. Furthermore, the instances of tweets relevant to the landslide disaster are dwarfed by irrelevant tweets, see \cref{fig:irrelevant}.

\begin{figure}[t]
	\centering
	\includegraphics[width=1\linewidth]{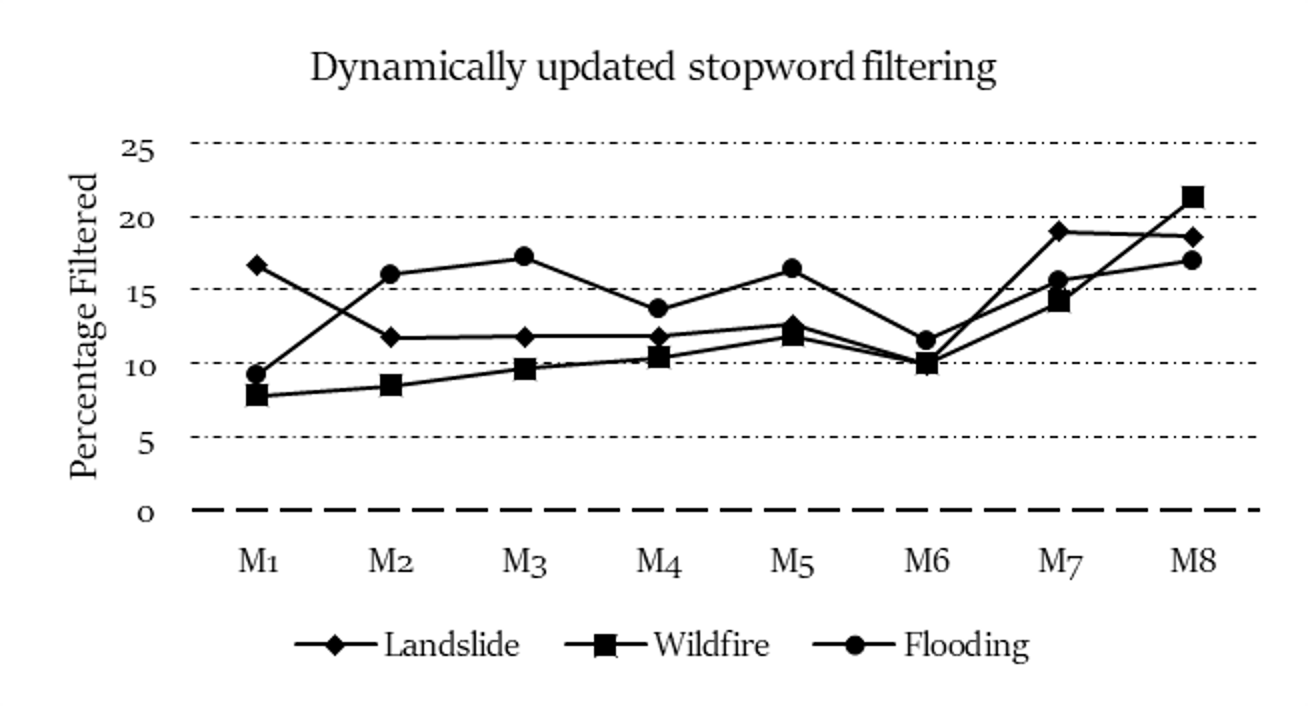}
	\caption{\textbf{Dynamic Stopword Filtering} Percentage irrelevant posts filtered by dynamically updated stopword list. Between 10-20\% of irrelevant items are filtered out this way. The remaining are procesed with more fine-tuned ML models.}
	\label{fig:stopwords}
\end{figure}

\PP{Data Cleaning} Probabilistic streams are noisy and need data cleaning before analysis. Cleaning can remove especially noisy examples. \system uses stopwords to perform data cleaning on probabilistic sources. In contrast to conventional approaches which create stopwords of common English terms, \system uses prior irrelevant data to continuously update the stopwords list. 

\system keeps track of irrelevant posts from probabilistic sources detected by the Event Detection module (see \cref{fig:system}. Periodically, most frequent terms in the irrelevant posts are added to the stopword list and old terms in the list are pruned. Specifically, the top-k most frequent terms in the irrelevant posts are added to the stopword list, replacing the prior frequent stopwords list; we let $k=5$. With dynamic data cleaning, \system is able to filter out between 10-20\% of irrelevant posts without requiring ML classifiers, see \cref{fig:stopwords}. This has an important advantage: earlier filtering reduces burden on the event processing system since irrelevant posts that are filtered out do not use up valuable compute resources.

\PP{Metadata Extraction} While traditional methods have used only the raw text for detection, more recent methods have exploited surrounding metadata to improve event detection or classification. For example \cite{drexel} uses metadata from users to improve sentiment classification, and \cite{credeye} uses metadata to determine user and information credibility. In \system, metadata extraction is similarly used to augment the raw text from the stream. Each event may require different metadata, and extraction is left to the framework deployer. In our landslide, flooding, and wildfire events, we primarily perform location extraction.

However, only 0.5\% of social media posts have a geo-tag, necessitating location extraction from the text content as well.	 Social sensor data has high noise, and Named Entity Recognition (NER), which is used for location extraction from text, often fails \cite{location} and misses many locations present in a post's text content. Because \system exploits the co-dependence between the corroborative sources and probabilistic sources, we use prior events detected in both dataflows to improve location extraction.

Prior events detected from the corroborative source dataflow have a location associated with them, since corroborative sources provide locations. The locations are then used for substring matches in the metadata extraction. As new events are detected, their locations augment the sub-string match list. Any locations matched with substring match are used to continuously train a NER using the corresponding probabilistic source text. In effect, our \system's applications combine three location extractors: off-the-shelf NER, continuously trained NER, and sub-string matches.


\PP{Corroborative Integration} Each real-world physical event has two primary attributes: location of event and time of occurrence. Since these spatio-temporal attributes are common to most physical event types we want to detect, \system takes advantage of these to further tune ML classifiers. Each application deployed on EventMapper has a collection of ML classifiers that take as input a text post with its metadata and provide as output a label of relevance or irrelevance (binary classification). Relevant posts are event classifications that will be shown to end users, and irrelevant posts are used for updating the stopword filters. However, as we showed in Related Work and \cref{fig:deterioration}, static ML classifiers exhibit performance deterioration over time. So, we need to update the ML classifiers over time.

\begin{figure}[t]
	\centering
	\includegraphics[width=1\linewidth]{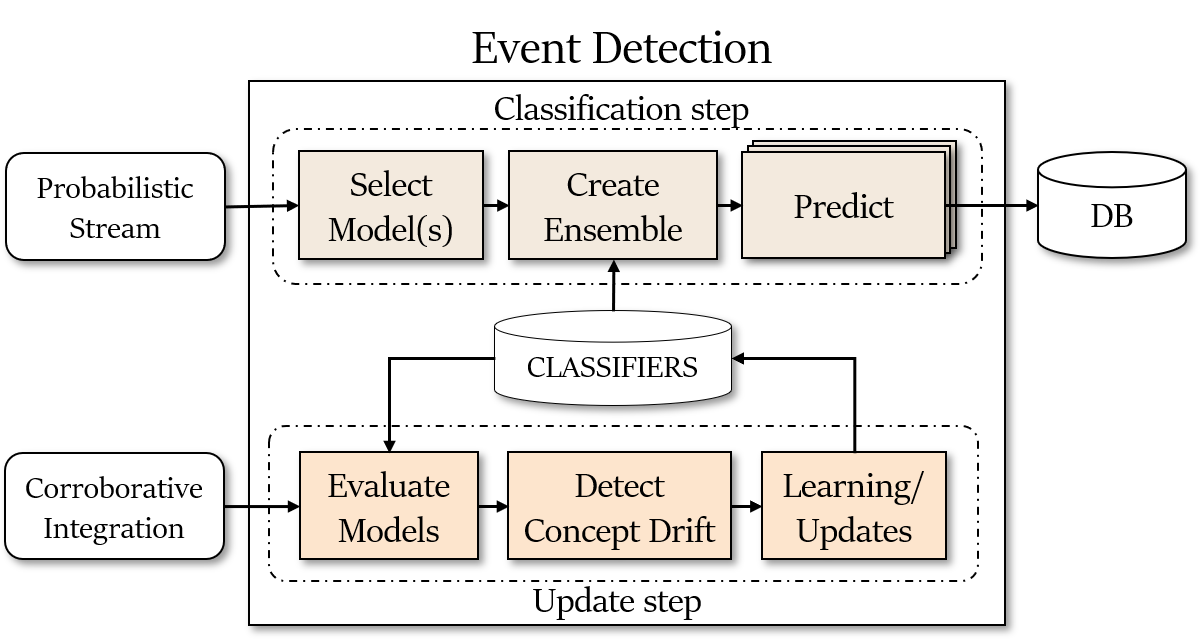}
	\caption{\textbf{Event Detection} We perform event detection using ML classifiers for probabilistic source posts that could not be labeled with corroborative integration. Fine-tuned classifiers are selected for each sample and a dynamic ensemble is constructed to label the sample. }
	\label{fig:eventdetection}
\end{figure}

Classifier updates require training data; recent approaches have used active learning to reduce the number of training data, and therefore, labeling cost. However, this is not scalable for streaming web data. \system solves this with \textit{Corroborative Integration}: by mapping both probabilistic posts with metadata and corroborative events to the same spatio-temporal grid, \system can automatically label some data points to create training data. If a corroborative event occurs at the same time and place as a probabilistic post with the event keywords, there is a high likelihood the social media post is relevant to the event. In effect, we perform weakly supervised labeling with the corroborative integration step. These weakly supervised labels can then be used as training data to update ML classifiers. While only 5\% of the probabilistic source data can be labeled with corroborative integration, it is 
sufficient to perform ML classifier updates, as we show in \textbf{Results}.

\PP{Negatively Labeled Data} One limitation of corroborative integration is only positive labels can be assigned to probabilistic posts. Negative labels cannot be assigned since probabilistic posts outside corroborative source coverage represent are of unknown relevancy because corroborative sources are delayed and have low coverage.

In \system, we use stopword filtered posts as negative samples. Since the stopwords within the text themselves present a strong signal to ML classifier and can skew it to perform the same function as a stopword filter, we remove the term from the post without replacement. So given a sentence of word tokens $S=\{w_1,w_2,s_1,w_3\}$ where $s_3$ is the stopword that triggered the stopword-filtering, we use the derived sentence $S=\{w_1,w_2,w_3\}$ as a negative label, removing $s_3$. During model updates or generation, positive and negative samples are converted to word embeddings with w2v for model training.

\subsection{Probabilistic Source Event Detection}
\PP{Event Detection} Since corroborative sources provide ground truth events, we focus on event detection in the probabilistic source dataflow. The Event Detection module, shown in \cref{fig:eventdetection}, performs two steps: (i) Update and (ii) Classification.

\PP{Update Step} \system uses the labeled data from corroborative integration to update existing classifiers. First, existing classifiers are evaluated on the labeled data with explicit drift detection using EDDM \cite{EDDM} (see Related Work). Classifiers with low performance can be pruned from the set of classifiers or flagged for update with the labeled data, based on user preferences. In our application implementations, we retain low performing models to archive them and use a copy of the model for updates. We then perform unsupervised concept drift detection to identify changes in the data distribution. We use the method from \cite{assed,mydrift} to detect drift using high density bands. \system maintains a memory of streaming points. We compare the distribution of the high density bands to the distribution of the streaming data using the Kullback-Leibler divergence. If drift is detected with the approach in \cite{mydrift}, \system uses the training data from corroborative integration to create new ML classifiers. For each updated or generated classifier, \system stores the classifier along with the training data for the classifier. The training data is indexed by the mean of the training data.

\PP{Classification Step} For each post from the probabilistic stream that could not be labeled with corroborative integration, \system classifies its relevance with the ML classifiers generated in the Update Step. While some approaches have used all stored models as an ensemble, \system dynamically creates an ensemble to best fit a data point. Given a probabilistic post (data point), \system sorts all classifiers on their training data's distance to the post. We measure distance using the train-ing data mean. The k-nearest classifiers are used to create an ensemble, with each classifier weighted by its distance to the data point. The dynamically created ensemble is used for relevancy prediction. 

\subsection{\system  Implementation}
We now describe the implementation for \system. The framework is designed for extensibility and allows us to deploy applications for different event detectors quickly. Each operation in the \system dataflow (\cref{fig:system}) is a process primitive. Instead of a linear dataflow with data passed between processes, \system uses a pub/sub interface to decouple process primitives. This accomplishes two things: (i) each process can be updated and managed independently, and (ii) the application remains fault-tolerant to crashes in any one process.

\PP{Streamers} EventMapper provides built-in streamers for probabilistic sources such as Twitter and Facebook. Our current work includes integration of YouTube and Instagram video and image streams as well. Streamers operate on user-defined, generic keywords for an event type provided as a configuration. In our applications, we use the following keywords for the streamers:

\squishitemize
\item Landslide: \textit{landslide}, \textit{mudslide}, \textit{rockslide}
\item Wildfire: \textit{wildfire}, \textit{brushfire}
\item Flooding: \textit{flood}, \textit{heavy rain}

\squishend

Since these are generic keywords, they include significant noise. Fine-tuning the keywords themselves can reduce our coverage and is not scalable, since drift may occur. So, we rely on \system’s fine-tuning to filter posts for event recognition.

\PP{Pub/Sub Interface}. \system uses a pub/sub inter-face to decouple processes. We use Apache Kafka as the pub/sub backend. Each process publishes its message to a Kafka topic, and subsequent processes susbcribe to the topic to receive the message. To ensure the same data point is not read twice during process crashes, we need to store read status for messages. Since Kafka is a minimal interface without control over read/write status of messages for expiration, we manage message read status with a Redis key value store. 

Each process in \system has an import key and export key, which are unique strings for each process. The import and export keys function both as pub/sub topics and Redis keys. An \system primitive process subscribes to its import key and publishes to its export key. Apache Kafka does not have support for recording message read/write status, so we use Redis to manage this.  For each message, \system updates on Redis the read/write values for its associated import key; it records message ID (a unique ID for each post), message offset, and partition. The offset and partition are used to recover an applicatio's position in the stream after a process crash. This also allows us to build many-to-many, one-to-many, and many-to-one dataflows in addition to the traditional one-to-one. Since each process manages its own import key’s offset with Redis, a message is guaranteed to be read onloy once by each process.

\PP{Process Management} \system manages each process to ensure continuous operation of an event detection application. To reduce overhead, each process is deployed independently and records its own process ID. \system checks process logs and process ID for non-operation, at which point any zombie executions are killed and the process is restarted. Subsequent failures in succession trigger an alert for the end-user to investigate process restart failures. 

\PP{Process Primitive} Each process in \system extends a base process primitive class. This allows applications to have flexibility in signal extraction for each module (data cleaning, metadata extraction, etc) while leaving message processing to \system. Therefore, while signal extraction logic needs to be written for each new event type, the process-to-process communication, fault-tolerance, and crash recovery are managed by \system.

\section{EventMapper Applications}

\label{sec:data}

In this section, we provide application details for our three desired events: landslides, wildfires, and flooding. We will cover results in the next section.

\subsection{Landslide Detection}
We select landslides as a desired event since they are a weak signal disaster with significant noise in social media streams; the use of the word \textit{landslide} is polysemous, meaning it carries multiple meanings. The word \textit{landslide} can refer to election landslides and a song \textit{Landslide} (by Fleetwood Mac) in addition to the disaster landslide.

\PP{Corroborative Sources} We use four corroborative sources for landslide detection:

\squishitemize
\item NASA TRMM provides landslide likelihood data in select locations around the globe. TRMM has three levels of predictions: 1-day, 3-day, and 7-day. Each prediction level provides: landslide likelihood using NASA's landslide models, closest location name, and latitude/ longitude. EventMapper uses  location name to update  substring match list for NER fine-tuning. We use the 1-day landslide predictions.
\item USGS Earthquake provides detected earthquakes around the globe. For each instance, it provides magnitude and latitude/longitude of the epicenter.
\item NOAA GHCND provides daily rainfall data at NOAA weather stations around the globe. Each station provides its latitude and longitude, along with rain in the past day. Due to a combination of old equipment, budget cuts, and progressive expansion, many stations do not provide up-to-date information.
\item News articles about landslides provide late cor-roboration since they are delayed by 2-3 days. EventMapper follows articles with an off-the-shelf API (NewsAPI) with the landslide disaster tag. Locations are extracted with NER since articles are long text and NER succeeds on the structured news text \cite{location,shorttext}.

\squishend

We use NASA TRMM and News as landslide ground-truth data, where available. We combine USGS and GHCND data, since heavy rainfall and earthquake in the same location indicates high probability of landslide \cite{litmus}, to provide secondary ground-truth data.

In the corroborative integration process, we map ground-truth events and probabilistic source posts to a spatio-temporal grid. For probabilistic posts at the same time and geographic location as ground-truth events, \system labels them as relevant posts.

The labeled posts from corroborative integration are used to fine-tune ML classifiers in the Event Detection process' Update Step. Each probabilistic post that could not be labeled with corroborative integration is processed with the Classification Step.

\subsection{Wildfire Detection}

Wildfires have flared up to a greater degree over the past two years due to climate change \cite{climate}. Furthermore, wildfires are expected to increase over the next years due to increased warming, longer fire seasons, increased emissions, and drier forests \cite{climate2,climate3}. Representative examples include the ongoing (at this time) Australian wildfires and the Amazon rainforest wildfires in 2018. Wildfires remain weak-signal however, since each wildfire instance is small and brush fires may crop up in isolated patches until it coalesces \cite{spotfire}. The term \textit{wildfire}, like \textit{landslide}, is also polysemous, since it can refer to pandemics and a Pokemon (children cartoon character).

\PP{Corroborative Sources} We use three corroborative sources for wildfire detection. The primary corroborative sources are from NASA's Fire Information for Resource Management System (FIRMS).
\squishitemize
\item FIRMS MODIS provides wildfire detections from satellite canopy data \cite{modis}. MODIS itself includes corroborative sources to clear its own false positives and improve fire detection.
\item FIRMS VIIRS \cite{viirs} is an evolution of MODIS ; while VIIRS has fewer spectral bands than MODIS, it can read higher resolution of fire data and is more sensitive to fire radiance. However, it is not fully deployed at the moment, so we use both MODIS and VIIRS.
\item News articles about wildfires provide late corroboration. \system uses NewsAPI to get articles with wildfire disaster tag. Locations are extracted with NER.

\squishend

MODIS and VIIRS provide ground truth events use to corroboratively label probabilistic source posts. Similar to landslide detection, labeled posts from corroborative integration fine-tune ML classifiers, which are in turn used for posts that could not be automatically labeled.

\subsection{Flooding Detection}

Similar to wildfires, flooding is expected to increase because of rising sea levels due to climate change \cite{flooding1}. There have already been increased river floods in the past few years \cite{flooding2}. The keywords for \textit{flooding} are also polysemous, since flood is used politically to refer to immigrants \cite{floodslur}, an antagonist in the \textit{Halo} video game series, and in reference to economics \cite{floodecon}.

\PP{Corroborative Sources} We use the following two sources for flooding detection:

\squishlist
\item USGS Flood Gauge provides flooding information across the US. Each instance is provided with latitude and longitude, along with flooding magnitude such as major, moderate, minor, or no flooding. Out of 9298 current gauges, 3427 gauges (37\%) are non-functional or not updated, reducing coverage. Also, the gauges are US-specific.
\item Similar to landslide and wildfires, we use NewsAPI to follow flooding tags to get late corroboration from news sources.
\squishend

The corroborative events are used for ML classifier fine-tuning. Classifiers are subsequently used for prediction.

\section{Results}

\begin{figure}[t]
	\centering
	\includegraphics[width=1\linewidth]{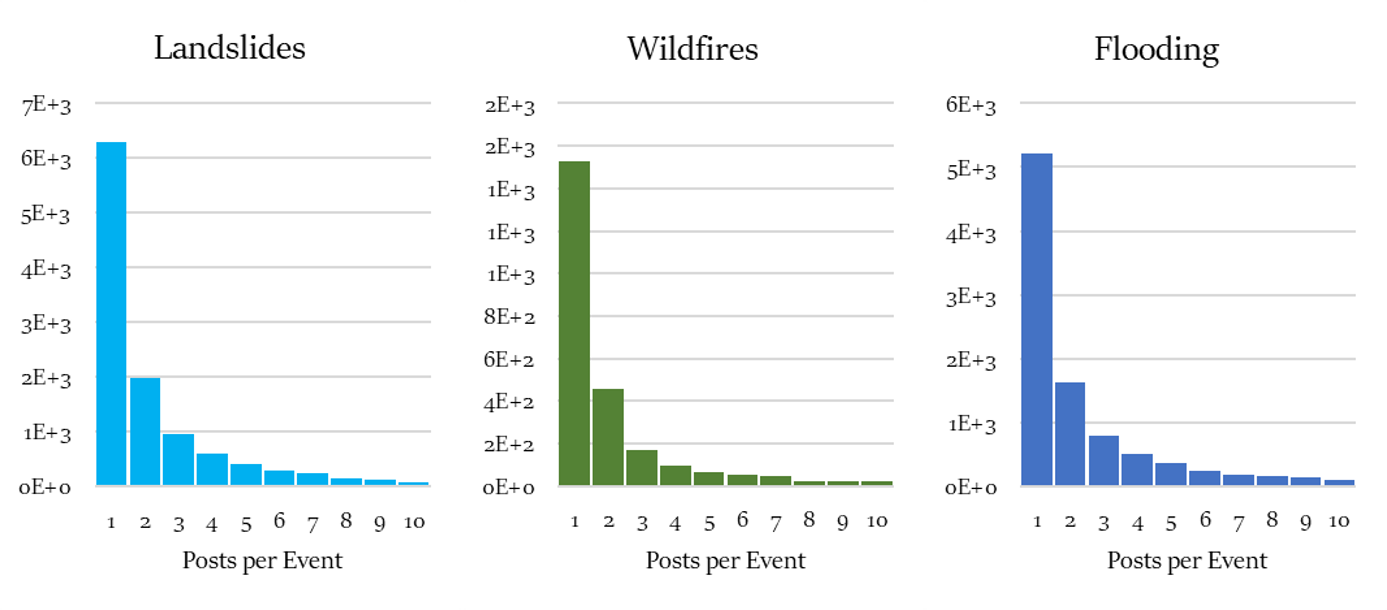}
	\caption{\textbf{Weak-Signal Events} Most events have 1-3 posts associated with them. The posts per event has a long tailed distribution; we show up to 10 posts per event for landslides, wildfires, and flooding events due to space.}
	\label{fig:weak}
\end{figure}

We evaluate our applications deployed on the \system  framework. We have providence evidence of drift-based performance deterioration in \cref{fig:deterioration}. We first describe the weak-signal nature of the events.

\begin{figure}[t]
	\centering
	\includegraphics[width=1\linewidth]{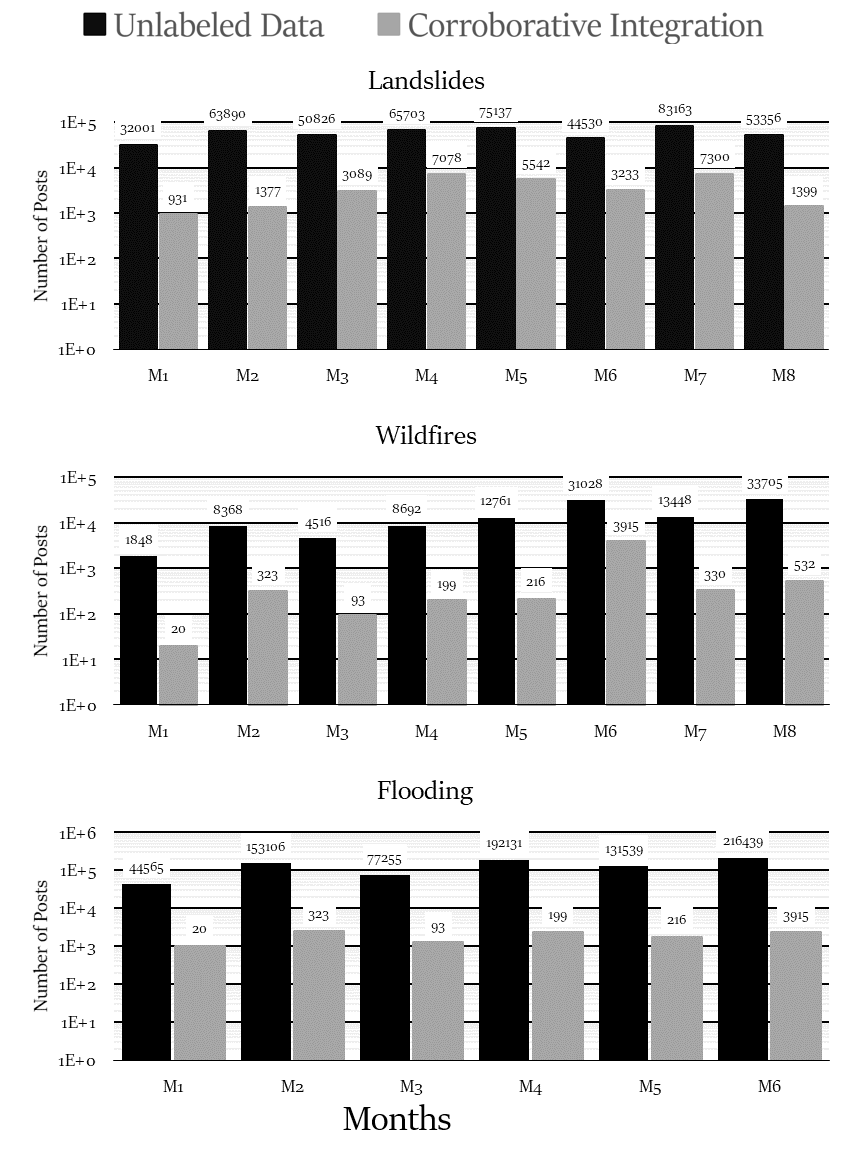}
	\caption{\textbf{Corroborative Integration} Less than 5\% of probabilistic source posts can be labeled with corroborative integration in each event type. We use these labeled posts for updating ML classifiers in the Classification Step of Event Detection. Y-Axis is log-scale.}
	\label{fig:hdi}
\end{figure}

\PP{Weak-Signal Events} As described in earlier, weak-signal events have few reports per event. Where corroborative event detection is not available, it is difficult to detect physical events from social media streams. We show evidence of the weak-signal nature of our events in \cref{fig:weak}. For each event, we record the number of posts present in the event. More than 50\% of events are associated with a single post, and 85\% of events have 5 or less posts associated with them. This presents challenges: an ML classifier must be robust to noise to avoid flagging false positives. However, this may reduce global coverage by ignoring weak-signal events that are hidden in noise or surrounding large-signal events.

\PP{Event Detection} Since \system continuously fine-tunes all stages of event recognition, from data cleaning (dynamic stopwords), metadata extraction (dynamic NER models), and classifiers (drift adaptive models), it is able to identify weak-signal events from 1-3 posts in addition to stronger-signal instances of weak-signal events (3\% of events have more than 15 posts per event). Integrating corroborative and probabilistic sources allows \system to take advantage of both source types: with ground-truth events from corroborative sources, \system can continuously fine-tune all stages of an event detection pipeline; simultaneously, \system ensures global coverage with probabilistic sources.

\begin{figure}[t]
	\centering
	\includegraphics[width=1\linewidth]{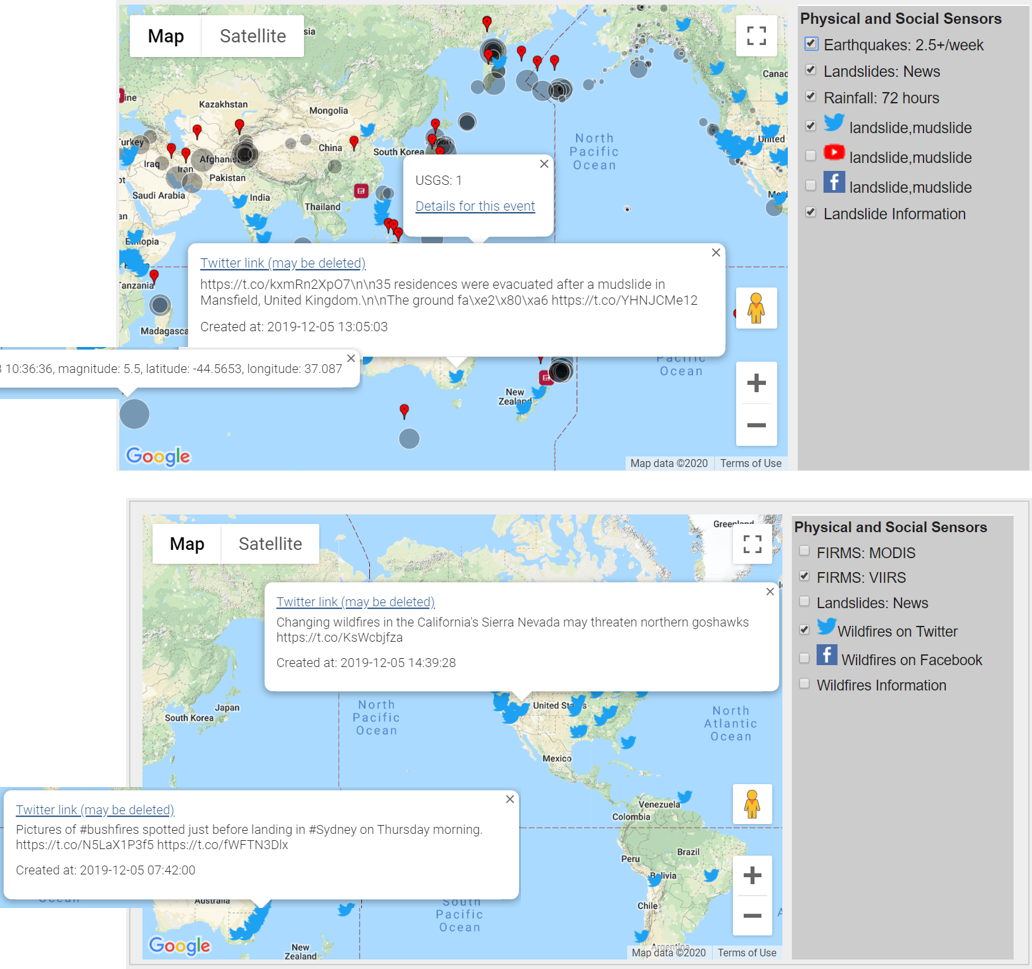}
	\caption{\textbf{Event Detection} Landslide and wildfire detection applications in \system showing detections from probabilistic source (Twitter) and corroborative integration (red markers). We omit flooding for space.}
	\label{fig:events}
\end{figure}

We show landslide and wildfire demo images in \cref{fig:events}. For landslides, we show map with events detected from Twitter, Facebook, News, and multi-source integration (red markers). For wildfires, we show detected events on Twitter.

\section{Conclusions}

In this paper, we have proposed the \system frame-work for event detection. The \system framework is designed for weak-signal event detection. It accomplishes this by integrating corroborative and probabilistic sources to exploit the co-dependence of event processing on each source type. With corroborative sources, \system can continuously fine-tune event processing on probabilistic sources. This allows for improved signal extraction for event detection, as we have shown: with continuous fine-tuning, we create a robust event detection pipeline that reduces long-term performance degradation. We also show \system applications' ability to detect weak-signal events. For each of our applications: landslides, wildfires, and flooding, \system identifies events from 1-3 tweets. Compared to work in \cite{sakaki2010earthquake,hurricane,gft}, which perform retroactive strong-signal detection on hundreds of thousands of posts, \system performs real-time event detection.
\pagebreak

\bibliographystyle{aaai}
\bibliography{main}

\begin{thebibliography}{}

\bibitem[\protect\citeauthoryear{Baena-Garc{\i}a \bgroup et al\mbox.\egroup
  }{2006}]{EDDM}
Baena-Garc{\i}a, M.; del Campo-{\'A}vila, J.; Fidalgo, R.; Bifet, A.; Gavalda,
  R.; and Morales-Bueno, R.
\newblock 2006.
\newblock Early drift detection method.
\newblock In {\em Fourth international workshop on knowledge discovery from
  data streams}, volume~6,  77--86.

\bibitem[\protect\citeauthoryear{Bevacqua \bgroup et al\mbox.\egroup
  }{2019}]{flooding2}
Bevacqua, E.; Maraun, D.; Vousdoukas, M.; Voukouvalas, E.; Vrac, M.; Mentaschi,
  L.; and Widmann, M.
\newblock 2019.
\newblock Higher probability of compound flooding from precipitation and storm
  surge in europe under anthropogenic climate change.
\newblock {\em Science advances} 5(9):eaaw5531.

\bibitem[\protect\citeauthoryear{Bifet and Gavalda}{2007}]{windowing}
Bifet, A., and Gavalda, R.
\newblock 2007.
\newblock Learning from time-changing data with adaptive windowing.
\newblock In {\em Proceedings of the 2007 SIAM international conference on data
  mining},  443--448.
\newblock SIAM.

\bibitem[\protect\citeauthoryear{Carter and Sutch}{2008}]{floodecon}
Carter, S.~B., and Sutch, R.
\newblock 2008.
\newblock Labor market flooding? migrant destination and wage change during
  america’s age of mass migration.
\newblock {\em Migration and development within and across borders: Research
  and policy perspectives on internal and international migration}  133--61.

\bibitem[\protect\citeauthoryear{Cervantes, Khokha, and
  Murray}{1995}]{floodslur}
Cervantes, N.; Khokha, S.; and Murray, B.
\newblock 1995.
\newblock Hate unleashed: Los angeles in the aftermath of proposition 187.
\newblock {\em Chicano-Latino L. Rev.} 17:1.

\bibitem[\protect\citeauthoryear{Chuanfei \bgroup et al\mbox.\egroup
  }{2010}]{stream}
Chuanfei, X.; Shukuan, L.; Lei, W.; and Jianzhong, Q.
\newblock 2010.
\newblock Complex event detection in probabilistic stream.
\newblock In {\em 2010 12th International Asia-Pacific Web Conference},
  361--363.
\newblock IEEE.

\bibitem[\protect\citeauthoryear{Cruz \bgroup et al\mbox.\egroup
  }{2012}]{spotfire}
Cruz, M.; Sullivan, A.; Gould, J.; Sims, N.; Bannister, A.; Hollis, J.; and
  Hurley, R.
\newblock 2012.
\newblock Anatomy of a catastrophic wildfire: the black saturday kilmore east
  fire in victoria, australia.
\newblock {\em Forest Ecology and Management} 284:269--285.

\bibitem[\protect\citeauthoryear{De~Albuquerque \bgroup et al\mbox.\egroup
  }{2015}]{coolpaper}
De~Albuquerque, J.~P.; Herfort, B.; Brenning, A.; and Zipf, A.
\newblock 2015.
\newblock A geographic approach for combining social media and authoritative
  data towards identifying useful information for disaster management.
\newblock {\em International Journal of Geographical Information Science}
  29(4):667--689.

\bibitem[\protect\citeauthoryear{Doornik}{2009}]{gft}
Doornik, J.~A.
\newblock 2009.
\newblock Improving the timeliness of data on influenza-like illnesses using
  google search data.
\newblock {\em Working paper}.

\bibitem[\protect\citeauthoryear{Eisenstein \bgroup et al\mbox.\egroup
  }{2014}]{lexical}
Eisenstein, J.; O'Connor, B.; Smith, N.~A.; and Xing, E.~P.
\newblock 2014.
\newblock Diffusion of lexical change in social media.
\newblock {\em PloS one} 9(11):e113114.

\bibitem[\protect\citeauthoryear{F{\"u}l{\"o}p \bgroup et al\mbox.\egroup
  }{2010}]{cepsummary}
F{\"u}l{\"o}p, L.~J.; T{\'o}th, G.; R{\'a}cz, R.; P{\'a}ncz{\'e}l, J.; Gergely,
  T.; Besz{\'e}des, A.; and Farkas, L.
\newblock 2010.
\newblock Survey on complex event processing and predictive analytics.
\newblock In {\em Proceedings of the Fifth Balkan Conference in Informatics},
  26--31.
\newblock Citeseer.

\bibitem[\protect\citeauthoryear{Gama \bgroup et al\mbox.\egroup
  }{2004}]{gamaB}
Gama, J.; Medas, P.; Castillo, G.; and Rodrigues, P.
\newblock 2004.
\newblock Learning with drift detection.
\newblock In {\em Brazilian symposium on artificial intelligence},  286--295.
\newblock Springer.

\bibitem[\protect\citeauthoryear{Gama \bgroup et al\mbox.\egroup
  }{2014}]{gamaA}
Gama, J.; {\v{Z}}liobait{\.e}, I.; Bifet, A.; Pechenizkiy, M.; and Bouchachia,
  A.
\newblock 2014.
\newblock A survey on concept drift adaptation.
\newblock {\em ACM computing surveys (CSUR)} 46(4):44.

\bibitem[\protect\citeauthoryear{Hemmings-Jarrett, Jarrett, and
  Blake}{2018}]{drexel}
Hemmings-Jarrett, K.; Jarrett, J.; and Blake, M.~B.
\newblock 2018.
\newblock (wksp) sentiment analysis of twitter samples that differentiates
  impact of user participation levels.
\newblock In {\em 2018 IEEE International Conference on Cognitive Computing
  (ICCC)},  65--72.
\newblock IEEE.

\bibitem[\protect\citeauthoryear{Kim}{2014}]{context}
Kim, Y.
\newblock 2014.
\newblock Convolutional neural networks for sentence classification.
\newblock {\em arXiv preprint arXiv:1408.5882}.

\bibitem[\protect\citeauthoryear{Kugler}{2016}]{gftfail3}
Kugler, L.
\newblock 2016.
\newblock What happens when big data blunders?

\bibitem[\protect\citeauthoryear{Lazer and Kennedy}{2015}]{gftfail2}
Lazer, D., and Kennedy, R.
\newblock 2015.
\newblock What we can learn from the epic failure of google flu trends. wired,
  october 2015.

\bibitem[\protect\citeauthoryear{Lee \bgroup et al\mbox.\egroup }{2006}]{viirs}
Lee, T.~E.; Miller, S.~D.; Turk, F.~J.; Schueler, C.; Julian, R.; Deyo, S.;
  Dills, P.; and Wang, S.
\newblock 2006.
\newblock The npoess viirs day/night visible sensor.
\newblock {\em Bulletin of the American Meteorological Society} 87(2):191--200.

\bibitem[\protect\citeauthoryear{Levin and Heimowitz}{2012}]{modis}
Levin, N., and Heimowitz, A.
\newblock 2012.
\newblock Mapping spatial and temporal patterns of mediterranean wildfires from
  modis.
\newblock {\em Remote Sensing of Environment} 126:12--26.

\bibitem[\protect\citeauthoryear{Liu, Stanturf, and Goodrick}{2010}]{climate3}
Liu, Y.; Stanturf, J.; and Goodrick, S.
\newblock 2010.
\newblock Trends in global wildfire potential in a changing climate.
\newblock {\em Forest ecology and management} 259(4):685--697.

\bibitem[\protect\citeauthoryear{Losing, Hammer, and Wersing}{2016}]{samknn}
Losing, V.; Hammer, B.; and Wersing, H.
\newblock 2016.
\newblock Knn classifier with self adjusting memory for heterogeneous concept
  drift.
\newblock In {\em 2016 IEEE 16th international conference on data mining
  (ICDM)},  291--300.
\newblock IEEE.

\bibitem[\protect\citeauthoryear{Lum and Isaac}{2016}]{gftfail1}
Lum, K., and Isaac, W.
\newblock 2016.
\newblock To predict and serve?
\newblock {\em Significance} 13(5):14--19.

\bibitem[\protect\citeauthoryear{Middleton \bgroup et al\mbox.\egroup
  }{2018}]{location}
Middleton, S.~E.; Kordopatis-Zilos, G.; Papadopoulos, S.; and Kompatsiaris, Y.
\newblock 2018.
\newblock Location extraction from social media: Geoparsing, location
  disambiguation, and geotagging.
\newblock {\em ACM Transactions on Information Systems (TOIS)} 36(4):40.

\bibitem[\protect\citeauthoryear{Musaev, Wang, and Pu}{2014}]{litmus}
Musaev, A.; Wang, D.; and Pu, C.
\newblock 2014.
\newblock Litmus: Landslide detection by integrating multiple sources.
\newblock In {\em ISCRAM}.

\bibitem[\protect\citeauthoryear{Pedruco \bgroup et al\mbox.\egroup
  }{2018}]{flooding1}
Pedruco, P.; Szemis, J.; Brown, R.; Lett, R.; Ladson, A.; Kiem, A.; Chiew, F.;
  et~al.
\newblock 2018.
\newblock Assessing climate change impacts on rural flooding in victoria.
\newblock In {\em Hydrology and Water Resources Symposium (HWRS 2018): Water
  and Communities},  645.
\newblock Engineers Australia.

\bibitem[\protect\citeauthoryear{Popat \bgroup et al\mbox.\egroup
  }{2018}]{credeye}
Popat, K.; Mukherjee, S.; Str{\"o}tgen, J.; and Weikum, G.
\newblock 2018.
\newblock Credeye: A credibility lens for analyzing and explaining
  misinformation.
\newblock In {\em Companion Proceedings of the The Web Conference 2018},
  155--158.
\newblock International World Wide Web Conferences Steering Committee.

\bibitem[\protect\citeauthoryear{Ren \bgroup et al\mbox.\egroup }{2018}]{kme}
Ren, S.; Liao, B.; Zhu, W.; and Li, K.
\newblock 2018.
\newblock Knowledge-maximized ensemble algorithm for different types of concept
  drift.
\newblock {\em Information Sciences} 430:261--281.

\bibitem[\protect\citeauthoryear{Ritter \bgroup et al\mbox.\egroup
  }{2012}]{noisy}
Ritter, A.; Etzioni, O.; Clark, S.; et~al.
\newblock 2012.
\newblock Open domain event extraction from twitter.
\newblock In {\em Proceedings of the 18th ACM SIGKDD international conference
  on Knowledge discovery and data mining},  1104--1112.
\newblock ACM.

\bibitem[\protect\citeauthoryear{Rong}{2014}]{w2v}
Rong, X.
\newblock 2014.
\newblock word2vec parameter learning explained.
\newblock {\em arXiv preprint arXiv:1411.2738}.

\bibitem[\protect\citeauthoryear{Sakaki, Okazaki, and
  Matsuo}{2010}]{sakaki2010earthquake}
Sakaki, T.; Okazaki, M.; and Matsuo, Y.
\newblock 2010.
\newblock Earthquake shakes twitter users: real-time event detection by social
  sensors.
\newblock In {\em Proceedings of the 19th international conference on World
  wide web},  851--860.
\newblock ACM.

\bibitem[\protect\citeauthoryear{Sakamoto \bgroup et al\mbox.\egroup
  }{2015}]{cepdrift}
Sakamoto, Y.; Fukui, K.-I.; Gama, J.; Nicklas, D.; Moriyama, K.; and Numao, M.
\newblock 2015.
\newblock Concept drift detection with clustering via statistical change
  detection methods.
\newblock In {\em 2015 Seventh International Conference on Knowledge and
  Systems Engineering (KSE)},  37--42.
\newblock IEEE.

\bibitem[\protect\citeauthoryear{Scheirer \bgroup et al\mbox.\egroup
  }{2012}]{openset}
Scheirer, W.~J.; de~Rezende~Rocha, A.; Sapkota, A.; and Boult, T.~E.
\newblock 2012.
\newblock Toward open set recognition.
\newblock {\em IEEE transactions on pattern analysis and machine intelligence}
  35(7):1757--1772.

\bibitem[\protect\citeauthoryear{Schoennagel \bgroup et al\mbox.\egroup
  }{2017}]{climate2}
Schoennagel, T.; Balch, J.~K.; Brenkert-Smith, H.; Dennison, P.~E.; Harvey,
  B.~J.; Krawchuk, M.~A.; Mietkiewicz, N.; Morgan, P.; Moritz, M.~A.; Rasker,
  R.; et~al.
\newblock 2017.
\newblock Adapt to more wildfire in western north american forests as climate
  changes.
\newblock {\em Proceedings of the National Academy of Sciences}
  114(18):4582--4590.

\bibitem[\protect\citeauthoryear{Sriram \bgroup et al\mbox.\egroup
  }{2010}]{shorttext}
Sriram, B.; Fuhry, D.; Demir, E.; Ferhatosmanoglu, H.; and Demirbas, M.
\newblock 2010.
\newblock Short text classification in twitter to improve information
  filtering.
\newblock In {\em Proceedings of the 33rd international ACM SIGIR conference on
  Research and development in information retrieval},  841--842.
\newblock ACM.

\bibitem[\protect\citeauthoryear{Stevens-Rumann \bgroup et al\mbox.\egroup
  }{2018}]{climate}
Stevens-Rumann, C.~S.; Kemp, K.~B.; Higuera, P.~E.; Harvey, B.~J.; Rother,
  M.~T.; Donato, D.~C.; Morgan, P.; and Veblen, T.~T.
\newblock 2018.
\newblock Evidence for declining forest resilience to wildfires under climate
  change.
\newblock {\em Ecology Letters} 21(2):243--252.

\bibitem[\protect\citeauthoryear{Suprem and Pu}{2019}]{assed}
Suprem, A., and Pu, C.
\newblock 2019.
\newblock Assed: A framework for identifying physical events through adaptive
  social sensor data filtering.
\newblock In {\em Proceedings of the 13th ACM International Conference on
  Distributed and Event-based Systems},  115--126.

\bibitem[\protect\citeauthoryear{Suprem, Musaev, and
  Pu}{2019}]{weaksupervision}
Suprem, A.; Musaev, A.; and Pu, C.
\newblock 2019.
\newblock Concept drift adaptive physical event detection for social media
  streams.
\newblock In {\em World Congress on Services},  92--105.
\newblock Springer.

\bibitem[\protect\citeauthoryear{Suprem}{2019}]{mydrift}
Suprem, A.
\newblock 2019.
\newblock Concept drift detection and adaptation with weak supervision on
  streaming unlabeled data.
\newblock {\em arXiv preprint arXiv:1910.01064}.

\bibitem[\protect\citeauthoryear{Wang, Hovy, and Dredze}{2015}]{hurricane}
Wang, H.; Hovy, E.; and Dredze, M.
\newblock 2015.
\newblock The hurricane sandy twitter corpus.
\newblock In {\em Workshops at the twenty-ninth AAAI conference on artificial
  intelligence}.

\end{thebibliography}
\end{document}